\documentstyle[preprint,aps,prb,epsfig,float]{revtex}

\begin{document}

\onecolumn
\title{
Anderson lattice in CeNiSi$_2$ intermediate valence compound
}
\author{S. O. Hong, Y. S. Kwon \thanks{To whom correspondence should be addressed. 
E-mail: yskwon@skku.ac.kr}}
\address{BK21 Physics Research Division and Institute of Basic Science, 
SungKyunKwan University, Suwon 440-746, South Korea}

\author{M. H. Jung} 
\address{MST-NHMFL, Los Alamos National Laboratory, 
Los Alamos, New Mexico 87545}

\date{\today}
\maketitle

\begin{abstract}
We present experimental results of electrical resistivity, 
Hall coefficient, magnetic susceptibility, and specific heat 
for single crystals of Ce-based intervalent compound CeNiSi$_2$. 
The results show similar behaviors observed in Yb-based intervalent compounds 
and support recent thoery of the Anderson lattice, in which
the Fermi-liquid coherence is gloval over the whole lattice. 
There is a low-temperature scale $T_{coh} \sim$ 50 K 
for the onset of Fermi-liquid coherence,
in addition to a high-temperature scale $T_K^* \sim$ 150 K 
for the Kondo-lattice condensation. 
Therefore, we conclude that two energy scales are generic 
in intermediate valence compounds based on Ce 
where the orbital degeneracy is smaller 
and where the size of the $4f$ orbital is larger 
than those based on Yb.
\end{abstract}

\newpage


The recently accepted picture of periodic intermediate valence (IV) and heavy fermion (HF) compounds is that, in addition to high-temperature Kondo singlet, Fermi-liquid coherent state sets in at low temperatures \cite{Burdin00}. There have been such examples of Yb-based IV compounds. In YbXCu$_4$ (X=Ag, Cd, In, Mg, Tl, Zn), there is a slow crossover from low-temperature Fermi-liquid regime to high-temperature local-moment regime, correlating with low conduction electron density background \cite{Lawrence01}. Another relevant example is YbAl$_3$, in which both the slow crossover and two energy scales are observed \cite{Cornelius02}. Current theoretical and experimental surveys support that these effects are well described by considering the Anderson lattice model in the limit of low conduction electron density \cite{Lawrence01} \cite{Cornelius02} \cite{Millis87}.

In this paper, we report results for Ce-based IV compound CeNiSi$_2$ applicable to the Anderson lattice. There is a high-temperature scale $T_K^* \sim$ 150 K, which is typical of IV compounds with Kondo temperature $T_K$ greater than 500 K. A low-temperature scale in magnetic susceptibility, specific heat, and Hall effect is observed below $T_{coh} \sim$ 50 K, which is the onset temperature of coherent Fermi-liquid $T^2$ behavior in the resistivity. We suggest that two energy scales are common for IV materials including Ce- and Yb-based compounds and that those are discussed in terms of the theory of the Anderson lattice, rather than the Anderson impurity model or the crystal field effect.

The single crystals were grown by the Czochralski method using a tetra-arc furnace. The electrical resistivity and Hall coefficient measurements were performed by a standard four-probe dc technique. The magnetic susceptibility was measured using a SQUID magnetometer and the specific heat was measured by an adiabatic dc method using a PPMS system. The data of LaNiSi$_2$ were used as a standard with no magnetic moment of $4f$ electrons.

Figure 1 shows the magnetic susceptibility $\chi(T)$, measured in a field of 0.1T parallel $(H \parallel b)$ and perpendicular $(H \perp b)$ to the $b$ axis. Over all the temperature range, the in-plane susceptibility $\chi_{\perp b}(T)$ is much larger than the $b$-axis susceptibility $\chi_{\parallel b}(T)$, indicating the easy magnetization direction $\parallel b$. A broad maximum around 100K is a characteristic of valence fluctuating systems, as reported earlier \cite{Pecharsky91}. The data can be fitted by
$\chi = N(2.54 \mu_B)^2 \nu(T) / 3k_B(T+T_{sf})$
with  the fractional occupation of Ce$^{3+}$ and Ce$^{4+}$ states given as 
$\nu(T) = 6 / \{6 + \exp[E_{ex}/k_B(T+T_{sf})]\}$, 
where $E_{ex}$ is the energy difference between Ce$^{3+}$ and Ce$^{4+}$ states and $T_{sf}$ is an effective thermodynamic temperature \cite{6}. This fit gives us $E_{ex}/k_B$ = 757.7 K and $T_{sf}$ = 270.2 K for $H \parallel b$, and $E_{ex}/k_B$ = 470.0 K and $T_{sf}$ = 157.5 K for $H \perp b$. From these parameters, the temperature dependence of the Ce valence can be calculated independently for both directions, which is shown in the insets. The valence is changed from mostly Ce$^{3+}$ at high temperatures to mostly Ce$^{4+}$ at low temperatures.  The rapid rise at low temperatures below 30 K might originate from a small amount of free Ce$^{3+}$ ions stabilized on lattice defects or a trace of some paramagnetic impurities, as often observed in other valence fluctuating systems \cite{7}. However, since our samples are determined into a sufficiently high quality and measured into low-temperature anomalies at that temperature in the other properties, we believe that this low-temperature anomaly is intrinsic in CeNiSi$_2$ and is associated with the Fermi-liquid coherence.

The electrical resistivity $\rho(T)$ measured with the current parallel $(I \parallel b)$ and perpendicular $(I \perp b)$ to the $b$ axis is illustrated in Fig. 2 for the whole temperature region and in the inset for the low temperature data in a $T^2$ scale. The $b$-axis resistivity $\rho_{\parallel b}(T)$ is much larger than the in-plane resistivity $\rho_{\perp b}(T)$, giving the ratio $\rho_{\parallel b}(T) / \rho_{\perp b}(T) \simeq$ 2. There is a sharp drop below $T^*_K \sim$ 150 K, indicating the onset of Kondo lattice ground states. Kondo lattices are characterized by a Fermi-liquid behavior below their Kondo temperatures. This Fermi-liquid state gives rise to a $T^2$ dependent resistivity. $\rho(T)$ of CeNiSi$_2$ exhibits such a $T^2$ behavior given by $\rho(T)= \rho_o + A T^2$ up to $T_{coh} \sim$ 50 K with $A$ = 0.029 $\mu \Omega cm / K^2$ and $\rho_o$ = 154.6 $\mu \Omega cm$ for $I \parallel b$, and $A$ = 0.018 $\mu \Omega cm / K^2$ and $\rho_o$ = 85.1 $\mu \Omega cm$ for $I \perp b$.

The presence of the Fermi-liquid coherent state is also observed in the Hall coefficient $R_H(T)$. In Fig. 3, $R_H(T)$ shows a broad minimum around $T^*_K \sim$ 150 K, which can be understood in a manner analogous to other Ce-based compounds with changes in the carrier density and skew scattering from the Ce 4f electrons \cite{8}. Using a simple one-band model that allows for $n$-type carriers, we estimate the temperature dependence of electron density $n(T)$ from $R_H(T)$. As shown in the inset, $n(T)$ is less than 0.6 per formula unit. The high temperature data of $n(T)$ is almost temperature independent, while $n(T)$ below $T_{coh} \sim$ 50 K shows a rapid rise. This result suggests a possible change of the Fermi surface as a result of the Fermi-liquid coherence.

In Fig. 4, we plot the linear coefficient of magnetic contribution to the specific heat $C_m = C(\rm {CeNiSi}_2) - C(\rm {LaNiSi}_2)$. The high temperature anomaly at 70 K is associated with that observed in $\chi(T)$, as is typical of valence fluctuating systems. For CeNiSi$_2$, the linear Sommerfeld coefficient is estimated to be $\gamma \simeq$ 57 mJ/mol K$^2$. It is interesting that with varying the transition metals and the semimetallic elements (increasing the size of ions), the $\gamma$ value is changed (enhanced) to 348 mJ/mol K$^2$ for CePdSi$_2$ \cite{9} and then 1700 mJ/mol K$^2$ for CePtSi$_2$ \cite{10}, and to 108 mJ/mol K$^2$ for CeNiGe$_2$ \cite{11}. This result could be attributed to a strong hybridization between the localized 4f electrons and the conduction electrons due to the increase in the density of states at the Fermi level, and to a weakening of the interatomic interactions due to both the increase of the interatomic distance and a change in the chemical nature. With further decreasing temperature below 20 T, $C_m /T$ displays an upturn associated with the Fermi-liquid coherence. It is noteworthy that the magnetic entropy $S_m(T)$ shown in the inset is slowly changed with temperature. At 300 K, $S_m(T)$ is only 70

Our experimental results for CeNiSi$_2$ IV compound demonstrate two energy scales $T^*_K \sim$ 150 K and $T_{coh} \sim$ 50 K and slow crossover as well as low conduction electron density $n \sim$ 0.1 - 0.2 per formula unit. These findings are well understood in a manner analogous to other Yb-based IV compounds such as YbXCu$_4$ and YbAl$_3$, supporting the recent model of Anderson lattice. Therefore, we speculate that the presence of two energy scales are common even for Ce-based IV compounds in the limit of low conduction electron density.

\begin{center}
\begin{figure}
\epsfig{figure=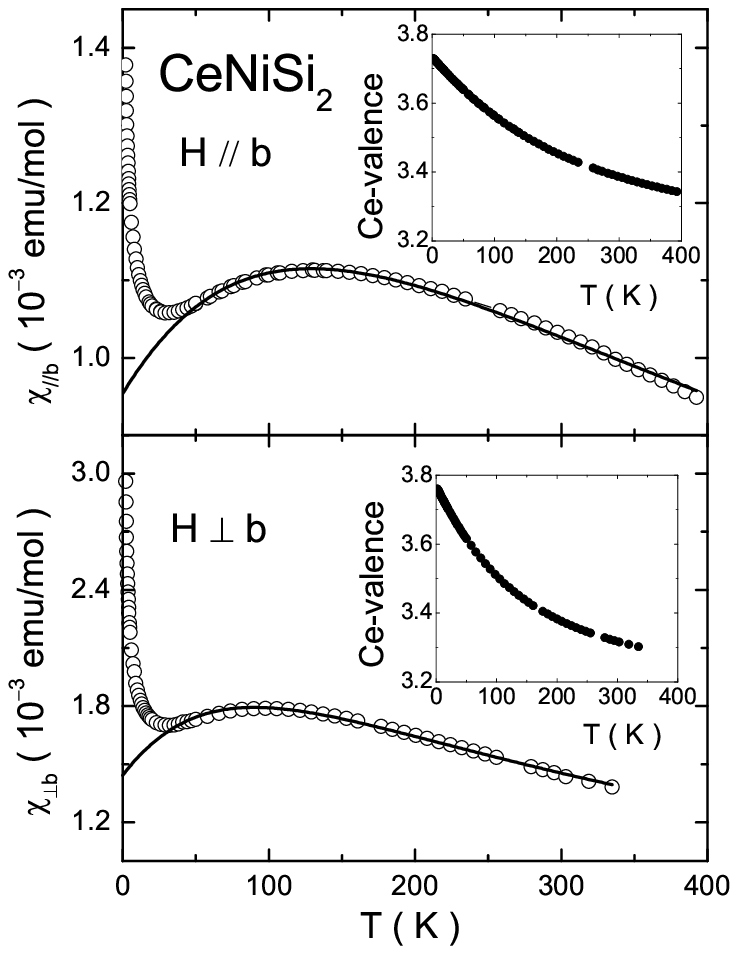,width=0.7\linewidth} 
\caption{Magnetic susceptibility $\chi(T)$ measured in a field of 0.1 T for $H \parallel b$ and $H \perp b$. The solid lines represent the best fits obtained with the model (see text) for each direction. Insets: the valence of Ce ion determined from the fitted parameters given in the text. \label{Fig_1}}
\end{figure}
\end{center}

\begin{center}
\begin{figure}
\epsfig{figure=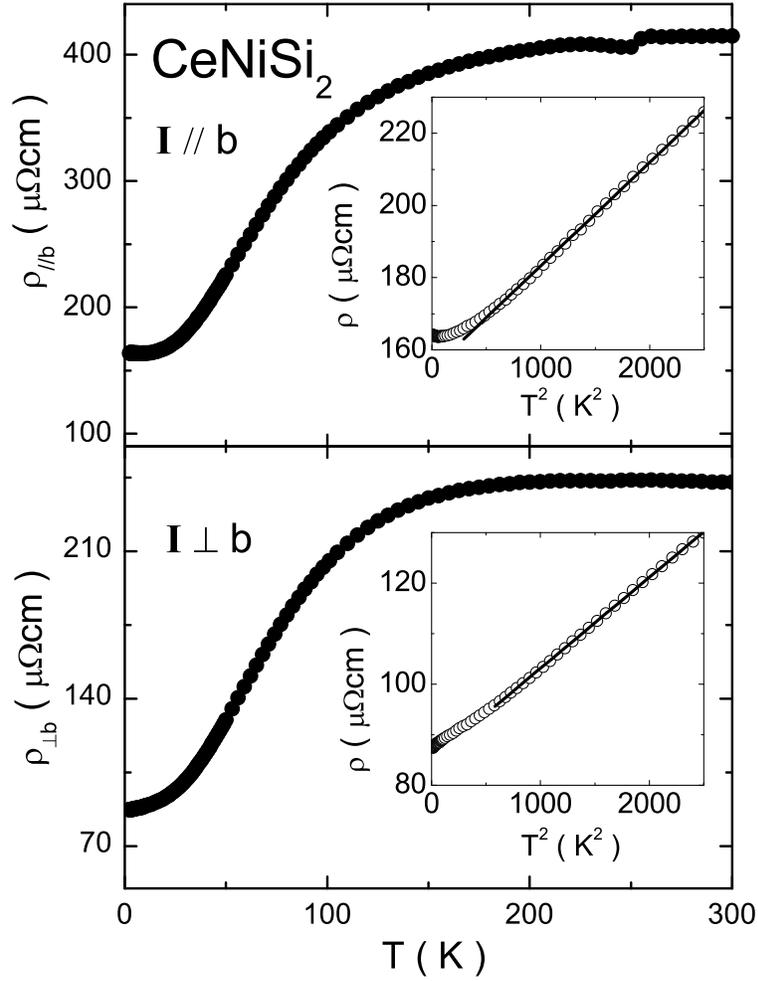,width=0.7\linewidth} 
\caption{Electrical resistivity $\rho(T)$ with the current $I \parallel b$ and $I \perp b$. Insets: $\rho(T)$ plotted in a $T^2$ scale. The solid lines show the Fermi-liquid $T^2$ behavior. \label{Fig_2}}
\end{figure}
\end{center}

\begin{center}
\begin{figure}
\epsfig{figure=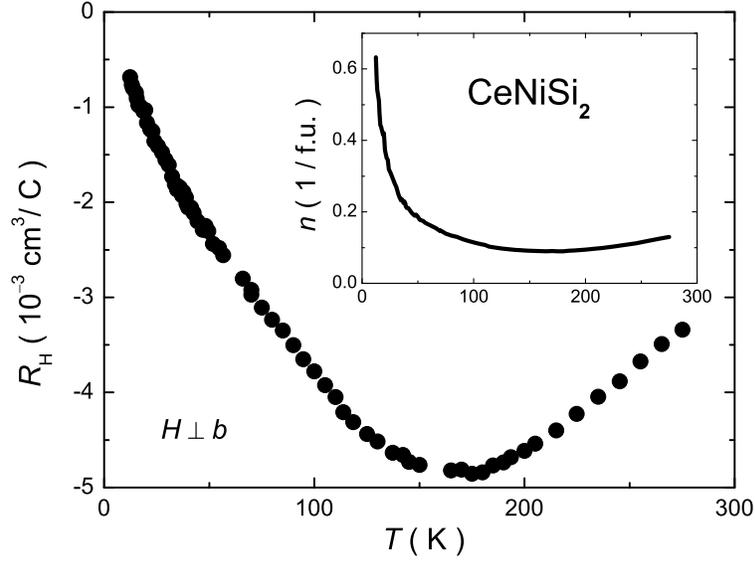,width=0.7\linewidth} 
\caption{Hall coefficient $R_H(T)$ measured in a field of 0.8 T for $H \perp b$. Inset: conduction electron density $n(T)$ estimated by using a simple one-band model. \label{Fig_3}}
\end{figure}
\end{center}

\begin{center}
\begin{figure}
\epsfig{figure=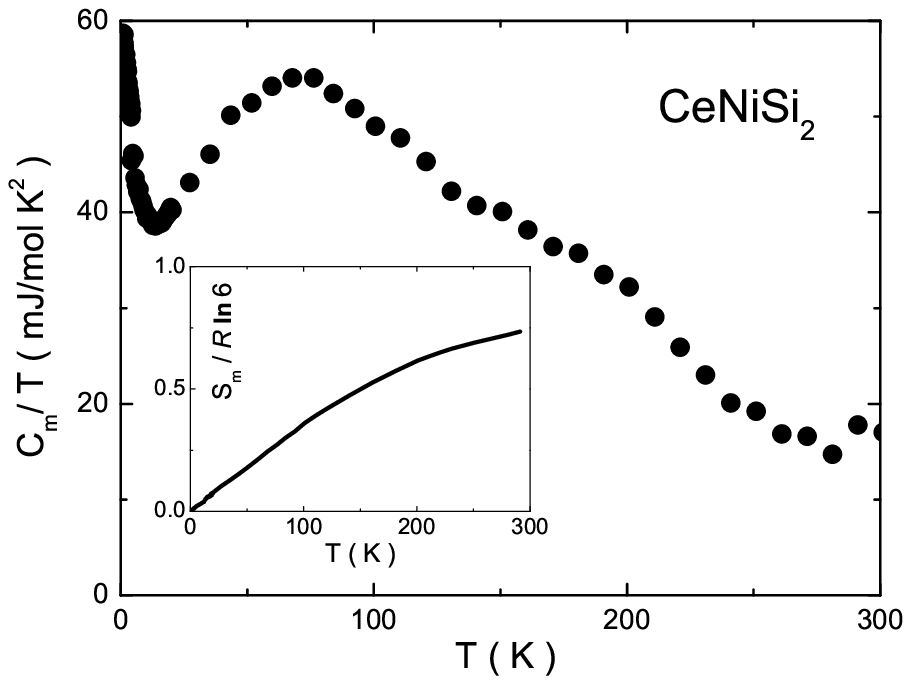,width=0.7\linewidth} 
\caption{Linear coefficient of magnetic contribution to the specific heat $C_m/T$. 
Inset: Magnetic entropy $S_m(T)$ normalized to the value of $R \ln6$. \label{Fig_4}}
\end{figure}
\end{center}


\begin{references}

\bibitem{Burdin00} 
S. Burdin, A. Georges, D. R. Grempel, Phys. Rev. Lett. {\bf 85} (2000) 1048.
\bibitem{Lawrence01} 
J. M. Lawrence, P. S. Riseborough, C. H. Booth, J. L. Sarrao, J. D. Thompson, R. Osborn, Phys. Rev. B {\bf 63} (2001) 05 4427.
\bibitem{Cornelius02} 
A. L. Cornelius, J. M. Lawrence, T. Ebihara, P. S. Riseborough, C. H. Booth, M. F. Hundley, P. G. Pagliuso, J. L. Sarrao, J. D. Thompson, M. H. Jung, A. H. Lacerda, G. H. Kwei, Phys. Rev. Lett. {\bf 88} (2002) 11 7201.
\bibitem{Millis87} 
A. J. Millis and P. A. Lee, Phys. Rev. B {\bf 35} (1987) 3394.
\bibitem{Pecharsky91}
V. K. Pecharsky, K. A. Gschneider, Jr., and L. L. Miller, Phys. Rev. B {\bf 43} (1991) 10906.
\bibitem{6}
B. C. Sales and D. K. Wolleben, Phys. Rev. Lett. {\bf 35} (1975) 1240.
\bibitem{7}
B. Chevalier, P. Rogl, K. Hiebel, and J. Etourneau, J. Solid State Chem. {\bf 107} (1993) 327.
\bibitem{8}
A. Fert and P. M. Levy, Phys. Rev. B {\bf 36} (1987) 1907.
\bibitem{9}
J. J. Lu, C. Tien, L. Y. Jang, and C. S. Wur, Physica B {\bf 305} (2001) 105.
\bibitem{10}
W. H. Lee, K. S. Kwan, P. Klavins, and R. N. Shelton, Phys. Rev. B {\bf 42} (1990) 6542.
\bibitem{11}
V. K. Pecharsky, K. A. Gschneidner, Jr., and C. L. Miller, Phys. Rev. B {\bf 43} (1991) 10906.

\end{references}
\end{document}